\documentclass[11pt]{article}
\usepackage{iap2000,epsfig,graphicx}

\def\be{\begin{equation}}
\def\ee{\end{equation}}
\def\bea{\begin{eqnarray}}
\def\eea{\end{eqnarray}}

\begin{document}
\title{CLUSTERING OF GALAXIES AND GROUPS IN THE NOG SAMPLE\\[16pt]}


\author{ G. GIURICIN, S. SAMUROVI\'C, M. GIRARDI, 
M. MEZZETTI, C. MARINONI}

\address{Department of Astronomy, Trieste Univ., via G. B. Tiepolo 
11,\\
34131 Trieste, Italy}

\maketitle\abstracts{
We use the two-point correlation function in redshift space, 
$\xi(s)$, 
to study the clustering of the galaxies and groups of the Nearby 
Optical 
Galaxy (NOG) Sample, which is a nearly all-sky, complete, 
magnitude-limited sample of $\sim$7000 bright and nearby optical 
galaxies. 
The correlation function of galaxies is well-described by a 
power-law, 
$\xi(s)= (s/s_0)^{-\gamma}$, with $\gamma\sim1.5$ and $s_0\sim 6.4\;
h^{-1}$ Mpc. We find evidence of morphological segregation between 
early- and 
late-type galaxies, with a gradual decreasing of the strength of 
clustering 
from the S0 to the late-type spirals, on intermediate scales. 
Furthermore, luminous galaxies (with $M_B\leq -19.5 + 5 \log h$)
are more clustered than dim galaxies. The groups show an excess 
of clustering with respect to galaxies. Groups with greater  
velocity dispersions, sizes, and masses are more clustered than those 
with lower values of these quantities.}

\section{Introduction}
Following  a series of studies (Marinoni {\it et al.}~\cite {ma98}, 
Marinoni
{\it et al.}~\cite{ma99}, Giuricin {\it et al.}~\cite{gi00a}; 
Marinoni 
{\it et al.}~\cite{ma00}) in which we investigate 
on the large-scale galaxy distribution in the nearby 
universe by using a nearly all-sky sample of optical galaxies, 
in this paper we analyze the clustering of the galaxies and groups 
of the Nearby Optical Galaxy (NOG) sample (Giuricin {\it et 
al.}~\cite{gi00a}).

The NOG is a complete, distance-limited ($cz\leq 6000$ km/s) and 
magnitude-limited 
($B\leq 14$ mag) sample of $\sim 7000$ 
nearby and bright optical galaxies, which covers $\sim2/3$ 
($|b|>20^{\circ}$) 
of the sky (8.27 sr). The degree of redshift completeness of the NOG 
is estimated 
to be 97\%.  The $B$-magnitudes are homogenized total blue magnitudes 
transformed to the standard system of the RC3 catalog (de Vaucouleurs 
{\it et al.}~\cite{de91}) and fully corrected for Galactic 
extinction, internal extinction, and 
K-dimming. Groups have been identified within the NOG by means of the 
hierarchical (H) 
and percolation (P) algorithms. About 45\% of the NOG galaxies were 
found to 
be group members. 

In this paper we use  the redshift-space two-point correlation 
function 
to analyze the clustering 
of the NOG galaxies (7028 galaxies with $cz\geq$ 50 km/s), the NOG H 
groups 
(474 groups with at least three members), and the NOG P groups (506 
groups  
with at least three members). Here we analyze  
the P groups obtained with the variant of the P algorithm in which 
the distance link parameter, corresponding to a minimum 
number density contrast of 80, is scaled with distance (it is 
$0.67\;h^{-1}$ Mpc at 4000 km/s), whilst the velocity link parameter 
is 
kept constant at the value of 350 km/s (here we adopt $H_0 = 100\; h$ 
km 
s$^{-1}$ Mpc$^{-1}$). The variant of the P algorithm in which both 
link parameters are 
scaled with distance leads to very similar groups (see Giuricin {\it 
et al.}~\cite{gi00a}
for details on group selection).

Almost all NOG galaxies (98.7\%) have a morphological classification. 
We divide 
NOG galaxies into two broad morphological bins, 
early-type galaxies (E-S0, $T<-1.5$) and late-type galaxies (Sp, 
$T\geq 
-1.5$) and into six fine morphological bins, E ($T<-2.5$), S0 
($-2.5\leq 
T < -1.5$), S0/a ($-1.5\leq T <0.5$), Sa ($0.5 \leq T < 2.5$), Sb 
($2.5 \leq 
T < 3.5$), and later types (hereafter denoted as Scd). The earliest 
bin, 
hereafter denoted simply as E, comprise also lenticulars, since it 
contains 
also objects broadly classified as E-S0.

\section{Calculating the Two-point Correlation Function}
We calculate the two-point correlation function in redshift space, 
$\xi(s)$, 
using the estimator of Hamilton~\cite{ha93}. We generate the random 
sample by filling 
the sample volume with a random distribution of the same number of 
points 
as in the data. The random points are distributed in depth according 
to the 
sample's selection function $S(s)$ which expresses the fraction of 
objects 
that are expected to satisfy sample's selection criteria.   

For magnitude-limited samples we calculate the weighted 
correlation function by replacing the counts 
of pairs with the weighted sum of pairs, $\sum w_i w_j$, which takes 
into 
account the selection effects acting on the sample used. The 
weighting scheme 
we adopt is that of equally weighted volumes, $w_i = 1/S(s_i)$.  
 
We calculate the selection functions of the whole sample of galaxies 
and 
of specific morphological subsamples in terms of their Schechter-type 
luminosity functions that we derive using redshifts as distance 
indicators 
(see Marinoni {\it et al.}~\cite{ma99} for details). 

As for groups, we have verified that the redshift distribution of 
the relatively rich groups are shifted to smaller values 
than that of galaxies. However, the redshift distributions of  
the magnitude-limited samples of P and H groups with at least 
three members are not significantly 
different from that of galaxies. Therefore, we use the conservative 
approach of computing their $\xi$(s) by assuming the same selection 
function adopted for galaxies. The same assumption was made by 
Ramella {\it et al.}~\cite{ra90}, Trasarti-Battistoni {\it et 
al.}~\cite{tr97}
and Girardi {\it et al.}~\cite{gbd00} in the calculation of the group 
correlation 
function.

A different approach used sometimes in the literature (e.g., Carlberg 
{\it et al.}~\cite{ca00}; Merch\'an {\it et al.}~\cite{me00}) to 
generate 
a random distribution of groups is to follow directly the observed 
redshift 
distribution of groups as if they were unclustered. However, this 
approach 
is conceptually questionable, since the results are sensitive to the 
clustering pattern that is present.

Interestingly, the direct derivation of the selection function of 
haloes hosting the whole hierarchical sequence from single galaxies 
to 
clusters (the catalogs of all galactic systems extracted from the 
NOG) 
is recently discussed by Marinoni {\it et al.}~\cite{ma00}.  
In an upcoming paper we will use this statistics to infer the 
correlation 
function of the halo distribution.

Moreover, we analyze volume-limited samples, which by definition 
contain 
objects that are luminous enough to be included in the sample when 
placed at the cutoff distance. This leads to uniformly selected data 
sets in which the same weight is assigned to each object.
 
Specifically, we extract volume-limited samples of galaxies at 
different 
depths. For instance, the volume-limited samples at depths of 6000 
(4000) 
km/s contain 2258 (1895) galaxies which are brighter than the 
magnitude limits of 
$M_B=-19.89 (-19.01) + 5 \log h$.
 
We further construct volume-limited samples of P and H groups with 
depth of 4000 km/s, by using suitably modified versions of the H and 
P 
algorithms in which the selection parameters which scale with 
distance are 
kept fixed at the values corresponding to 4000 km/s. These 
volume-limited 
samples contain 140 H and 141 P groups with at least three members.

We average the results obtained using 
many different replicas of the random sample (in general 50 replicas 
in the 
case of galaxies and 400 in the case of groups). 

We calculate the errors for the correlation function by using 100 
bootstrap resamplings of the data (e.g., Ling {\it et 
al.}~\cite{li86}). 

Since, in general, $\xi(s)$ is well described by a power law over a 
fairly large interval of $s$, we always fit $\xi(s)$ to the form 
$\xi(s) = (s/s_0)^{-\gamma}$ with a non-linear weighted least-squares 
method in the intervals where $\xi(s)$ is reasonably fitted by a 
single 
power law. 

\section{The Correlation Function of NOG galaxies}

\subsection{Results from the whole NOG}
Fig. 1 shows the redshift-space correlation function from the 
whole NOG (1 $\sigma$ error bars are shown). From a power-law 
fit calculated in the interval 
$2.7 - 12\;h^{-1}$ Mpc, we find a correlation length of $s_0=6.42 \pm 
0.11\;h^{-1}$ Mpc and a slope $\gamma= 1.46 \pm 0.05$. On small 
scales ($s < 2\;h^{-1}$ Mpc) $\xi(s)$ tends to flatten because of the 
effects of 
peculiar motions, whereas on large scales ($s > 15 \;h^{-1}$ Mpc) 
it tends to steepen.

\begin{figure}
\psfig{figure=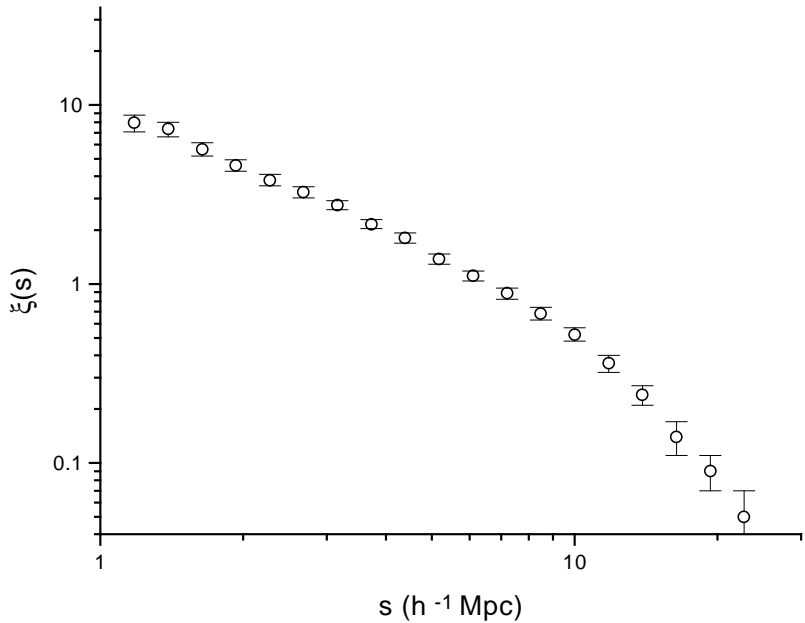,angle=270,height=10cm}
\caption{The redshift-space galaxy correlation function from the 
whole NOG; 1$\sigma$  error bars are shown.
\label{fig:corrfunone}}
\end{figure}

The NOG $\xi(s)$ is in good agreement with 
the results of most redshift surveys of optical galaxies 
(especially the LCRS, see Tucker {\it et al.}~\cite{tu97}), 
which are characterized by very different geometries, volumes and 
selection 
criteria (see, e.g., Willmer {\it et al.}~\cite{wi98} and references 
cited therein). We derive a smoother $\xi(s)$ than previous works 
probing larger volumes (see, e.g., the results coming from the 
Stromlo-APM (Loveday {\it et al.}~\cite{lo95}) and Durham-UKST 
(Ratcliffe {\it et al.}~\cite{ra96}) redshift surveys), because the 
NOG 
contains a larger number of galaxies.

The agreement between different galaxy correlation functions derived 
for a wide range of volumes and sample radii is in contrast with the 
fractal interpretation of the galaxy distribution in the universe.

\subsection{Morphological segregation}
Subdividing the NOG into several morphological types, we note 
a pronounced morphological segregation between the E-S0 galaxies 
($N=1036$), characterized by $s_0\sim 11.1\pm0.5\; h^{-1}$ Mpc and 
$\gamma\sim 1.5\pm0.1$, and the Sp objects ($N=5899$), characterized 
by $s_0\sim 5.6\pm0.1\;h^{-1}$ Mpc and $\gamma\sim 1.5\pm0.1$). 
More specifically, there is a gradual increase of 
the strength of clustering from the Scd to the S0 objects, especially 
on intermediate scales, but this tendency does not extend to the 
earliest types which do not show a greater degree of clustering 
than the S0s (see Figs 2 and 3).

\begin{figure}
\psfig{figure=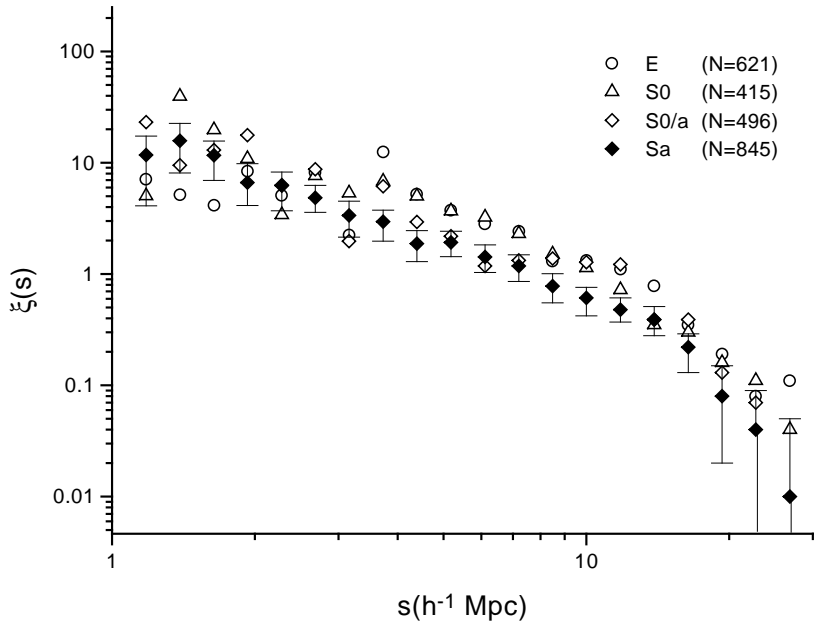,angle=270,height=10cm}
\caption{Comparison of the correlation functions for the E, S0, S0/a, 
and Sa morphological types. 
For the sake of clarity, 1$\sigma$ error bars are shown for the last 
sample only. The number of objects is indicated.
\label{fig:corrfuntwo}}
\end{figure}

\begin{figure}
\psfig{figure=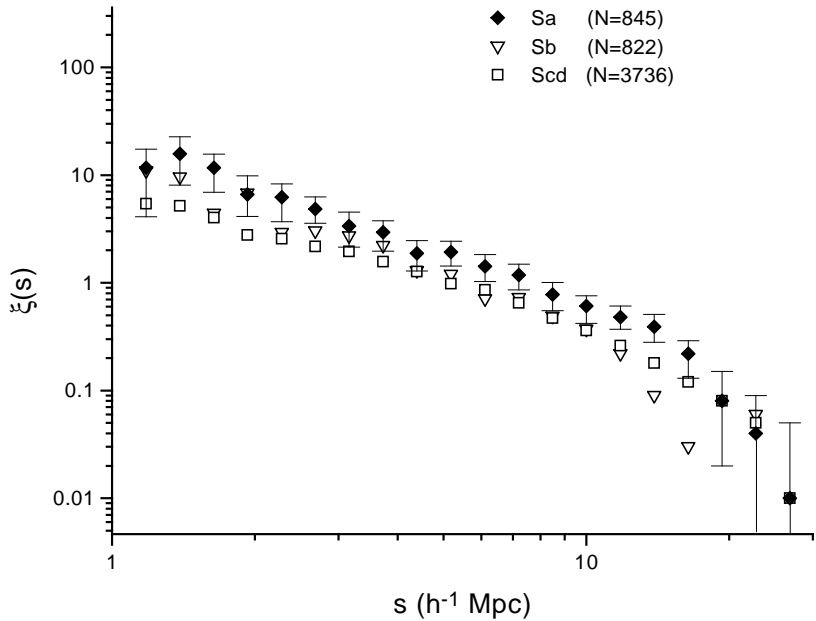,angle=270,height=10cm}
\caption{Comparison of the correlation functions for the  Sa, Sb, and 
Scd morphological types. For the sake of clarity,  1$\sigma$ error 
bars are shown for the first sample only. The number of objects is 
indicated.
\label{fig:corrfuntri}}
\end{figure}

Contrary to some recent claims (Hermit {\it et al.}~\cite{he96}; 
Willmer {\it et al.}~\cite{wi98}) the relative bias factor 
($\sim$ 1.7) between early- and late-type objects appears to be 
constant with scale. 

\subsection{Luminosity segregation}
We analyze different volume-limited samples of galaxies at 
different depths to search for luminosity 
segregation within a given sample. We find that the luminous 
galaxies (both  early and late types) are more clustered 
than the dim objects. The luminosity segregation 
starts to become appreciable only at relatively 
high luminosities ($M_B\leq -19.5 + 5\log h$, i.e. $L\geq 0.6 L^*$) 
and is independent on scale (at least up to $10\;h^{-1}$ Mpc).
Our results are in line with a series of papers which reported 
evidence of luminosity segregation, although there is little 
consensus in the literature about the range of morphological 
types and luminosities at which the effect occurs (e.g., cf. 
Loveday {\it et al.}~\cite{lo95} and Willmer {\it et 
al.}~\cite{wi98}). 

The very luminous galaxies ($M_B\leq -21 + 5 \log h$; $L\geq 
2.4\;L^*$) 
reside preferentially in binaries and groups (though not in clusters) 
and are characterized by $s_0\sim 12\;h^{-1}$ Mpc. 

The morphological and luminosity segregations appear to be two 
separate effects. The fact that they are detected also on large 
scales 
favors the interpretation that, on scales greater than $\sim 1$ Mpc, 
the bulk of these effects is likely to be mostly primordial in 
origin, i.e., inherent in schemes of biased galaxy formation 
(e.g., Bardeen {\it et al.}~\cite{ba86}) and not induced by late 
environmental effects.
 
\section{The correlation function of NOG groups}
The NOG P and H groups (with at least three members) are in general 
poor systems with typical (median) internal velocity dispersion
of $\sim100$ km/s, a virial radius of $\sim 0.8\;h^{-1}$ Mpc, 
a virial mass of $7\cdot 10^{12}\;h^{-1}\;M_{\odot}$, with the former 
groups
being, on average, slightly more massive and smaller in size than the 
latter 
ones (see, e.g., Tucker {\it et al.}~\cite{tu00} and Girardi \& 
Giuricin~\cite{gg00}  
for an overview of the properties of other catalogues of groups).

In Fig. 4 we show the weighted correlation functions of the H groups 
and P groups (with at least three members) compared to that 
of all galaxies. On small scales ($<3.5\;h^{-1}$ Mpc), the group 
correlation functions start dropping because of the anti-correlation 
due to the typical size of groups.

\begin{figure}
\psfig{figure=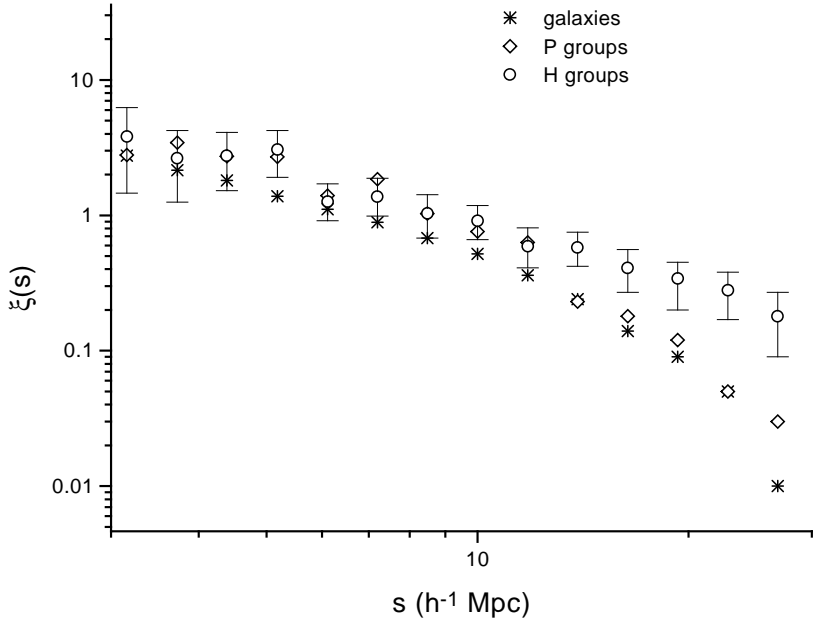,angle=270,height=10cm}
\caption{Comparison of the correlation functions for the galaxies, P 
groups, and H groups; 1$\sigma$ error bars are shown for the H groups 
only.
\label{fig:corrfunfour}}
\end{figure}
The correlation functions of the P and H groups show greater 
amplitudes than 
that of galaxies with an excess of clustering by a factor $\sim1.5$ 
and 
$\sim2$, respectively. Although the two samples of groups are 
significantly 
different in the distributions of the above-mentioned dynamical 
quantities, 
they show similar clustering properties, 
at least on intermediate scales ($s<10\;h^{-1}$ Mpc).   

Power-law fits over the interval $3.5 \leq s \leq 
20\;h^{-1}$ Mpc give $s_0=7.8 \pm 0.7\;h^{-1}$ Mpc, $\gamma = 2.0 \pm 
0.2$ 
and $s_0 = 8.4 \pm 0.7\;h^{-1}$ Mpc, $\gamma = 1.3 \pm 0.2$, 
respectively.
Thus, groups appear to have a degree of clustering intermediate 
between galaxies and clusters. 
Our results are in good agreement with those by Girardi 
{\it et al.}~\cite{gbd00}, who reported a similar excess of 
clustering 
for the groups identified in the CfA2 and SSRS2 redshift surveys. 
(see Girardi {\it et al.}~\cite{gbd00} for  earlier controversial 
results on the group correlation function).

We use the volume-limited samples of P and H groups at the depth of 
4000 
km/s to explore the dependence of the strength of clustering on some 
properties of groups. We find that groups with greater 
internal velocity dispersions, virial radii, mean pairwise member 
separations, and virial masses, tend to be more clustered than those 
with lower values of these quantities. On the other hand, there is 
no difference in the degree of clustering between groups with small 
and large proportions of early-type galaxies and with long and short 
crossing times.   
 
Further details on the clustering analysis of the NOG sample will 
be presented in Giuricin {\it et al.}~\cite{gi00b}. Work on 
redshift-space distortions in the NOG is in progress.


\section*{References}
\begin{thebibliography}{99}
\bibitem{ba86} Bardeen, J. M., Bond, J. R., Kaiser, N. Szalay, A. S. 
1986,
               ApJ, 304, 15. 
\bibitem{ca00} Carlberg, R. G. {\it et al.} 2000, ApJ, in press 
(astro-ph/0008201).
\bibitem{de91} de Vaucouleurs, G. {\it et al.} 1991, Third Reference 
Catalogue of Bright 
               Galaxies (New York: Springer).
\bibitem{gbd00} Girardi, M., Boschin, W., da Costa, L. N. 2000, A \& 
A, 353, 57. 
\bibitem{gg00}  Girardi, M. \& Giuricin, G. 2000, ApJ, 540, 45.
\bibitem{gi00a} Giuricin, G., Marinoni, C., Ceriani, L., Pisani, A. 
2000a, ApJ, in press 
                (preprint astro-ph/0001140).
\bibitem{gi00b} Giuricin, G., Samurovi\'c, S., Girardi, M., Mezzetti, 
M., 
                Marinoni, C. 2000b, submitted.
\bibitem{ha93} Hamilton, A. J. S. 1993, ApJ, 417, 19.
\bibitem{he96} Hermit, S. {\it et al.} 1996, MNRAS, 283, 709. 
\bibitem{li86} Ling, E. N., Frenk, C. S., Barrow, J. D. 1986, MNRAS, 
223, 21P.
\bibitem{lo95} Loveday, J., Maddox, S. J., Efstathiou, G., Peterson, 
B. A. 1995, 
               ApJ, 442, 457.  
\bibitem{ma00} Marinoni, C., Hudson, M. J., Giuricin, G. 2000, in 
preparation.
\bibitem{ma98} Marinoni, C., Monaco, P., Giuricin, G., Costantini, B. 
1998, 
               ApJ, 505, 484. 
\bibitem{ma99} Marinoni, C., Monaco, P., Giuricin, G., Costantini, B. 
1999,
               ApJ, 521, 50.
\bibitem{me00} Merch\'an, M. E., Maia, M. A. G., Lambas, D. G. 2000, 
               ApJ, in press (preprint astro-ph/0006407).
\bibitem{ra90} Ramella, M., Geller, M. J., Huchra, J. P. 1990, ApJ, 
353, 51.
\bibitem{ra96} Ratcliffe, A. {\it et al.} 1996, MNRAS, 281, L47.
\bibitem{tr97} Trasarti-Battistoni, R., Invernizzi, G., Bonometto, S. 
A. 1997, 
               ApJ, 475, 1.
\bibitem{tu97} Tucker, D. L. {\it et al.} 1997, MNRAS, 285, L5.
\bibitem{tu00} Tucker, D. L. {\it et al.} 2000, ApJ, in press 
(preprint astro-ph/0006153).
\bibitem{wi98} Willmer, C. N. A., da Costa, L. N., Pellegrini, P. S. 
1998, 
               AJ, 115, 869.

\end{thebibliography}
\end{document}